# Manipulation of magnetization in Pd(100) ultrathin films with quantum well structure using modification of Schottky barrier potentials

Hidetake Tanabe,[1] Shunsuke Sakuragi,[2] and Tetsuya Sato[1]

[1]*Department of Applied Physics and Physico-Informatics, Keio University, 3-14-1 Hiyoshi, Kohoku-ku, Yokohama 223-0061, Japan*

[2] *Institute for Solid State Physics, University of Tokyo, 5-1-5 Kashiwanoha, Kashiwa 277-8581, Japan*

The magnetization of Pd(100) ultrathin films that show ferromagnetism due to quantum well states was manipulated by changing the quantum well state with an applied bias voltage. The voltage dependence of the magnetic moment of Pd/SrTiO$_{3-x}$/Ti/Au intrinsically depends on the Pd film thickness. The induced change in the magnetic moment is due to the modulation of the phase shift at the interface between the Pd thin film and the semiconductor SrTiO$_{3-x}$ substrate.

In recent years, there have been reports regarding manipulation of the magnetism through the application of an electric field with the goal to develop a low-energy technology in the field of spintronics or multiferroics.[1-4] However, clear guidelines to manipulate the magnetic moment using an external field have not been established because the underlying mechanism has not been sufficiently clarified. Controlling the magnetic properties of Pd has been studied, which exhibits ferromagnetism at the nanoscale despite being paramagnetic in bulk.[5] Pd(100) ultrathin films periodically show ferromagnetism depending on the film thickness.[6] The appearance of ferromagnetism in Pd(100) ultrathin films has been interpreted in terms of the Stoner criterion[7] and based on the electric state of Pd.[6,8] Therefore, changes in the magnetic moment, such as nonmagnetic/ferromagnetic switching, can be realized by suitably modifying the physical parameters that dictate the ferromagnetic behavior.

Quantum well (QW) states are formed in Pd (100) ultrathin films, which generate discretized energy levels. When the energy levels overlap with the Fermi surface at certain film thicknesses, the density of the states $D(\varepsilon)$ around the Fermi energy $\varepsilon_F$ increases. This can cause the appearance of ferromagnetism in Pd(100) ultrathin films based on the Stoner criterion for ferromagnetism[7] as shown in Eq. (1),

$$ID(\varepsilon_F) = 1, \quad (1)$$

where $I$ is the exchange integral.

The QWs in metal thin films studied to date[8-13] have been described using the phase accumulation model[14] as Eq. (2),

$$2\pi k_z N + \Phi = 2\pi j, \quad (2)$$

where $k_z$ is the wave vector perpendicular to the film plane in units of $2\pi/a$, $N$ is the number of atomic layers, $\Phi$ is a correction term for the scattering phase shift at the surface or interface, and $j$ is the integer quantum number. When $k_z = k_F$ ($k_F$ is the Fermi wave vector) is satisfied, the Pd(100) ultrathin films exhibit ferromagnetism with the increase of $D(\varepsilon_F)$. According to Eq. (2), periodic changes in the physical properties that depend on $N$ have been observed.[15,16] In addition, the QWs also depend on $\Phi$, which is determined by the potential at the surface or interface.[17] Therefore, the manipulation of the magnetization can be realized when the QWs are modified by changing the potential.

In this study, we changed the interface potential by focusing on the Schottky potential formed at the metal/semiconductor interface. It is known that the Schottky potential can vary with an applied voltage, so changes in $\Phi$ are expected by altering the interface potential with a bias voltage. Therefore, we manipulated the magnetization of Pd(100) ultrathin films using an applied voltage to change the QWs.

A Ti(60 nm)/Au(80 nm) electrode was deposited on the back side of the SrTiO$_3$ (STO) substrate to obtain an ohmic contact. The STO substrate was then transferred to a molecular beam epitaxy (MBE) chamber (~$3 \times 10^{-7}$ Pa) and oxygen vacancies were introduced into the substrate by annealing it for 3 h at 693 °C (SrTiO$_{3-x}$) to give the semiconductor characteristics.[18] The Pd(100) thin films were then deposited using an MBE apparatus according to the following three-step growth method.[6,19,20] First, one-fifth of the total thickness of the Pd film was deposited at 400 ℃. Second, the remaining Pd was deposited at room temperature. Third, the deposited film was heated to 420 ℃ for annealing. The quality and film thickness were checked using reflection high-energy electron diffraction (RHEED) and X-ray reflectivity measurement (XRR). The *I - V* measurements were performed to evaluate the Schottky characteristics using a KEITHLEY Model 2400 Series Source Meter. The magnetization measurements were performed using a superconducting quantum interference device (SQUID) magnetometer while applying various voltages. All magnetic measurement was performed at room temperature. A schematic drawing of the prepared film sample is shown in Fig. 1(a).

The RHEED image of the Pd/SrTiO$_{3-x}$ is shown in Fig. 1(b), where the observed streak pattern indicates the favorable orientation and flatness of the sample. Figures 1(c) and 1(d) show the *I-V* curves of the Pd/SrTiO$_{3-x}$/Ti/Au with a 3.57 (sample A) and 3.87 nm (sample B) film thickness of Pd layer. Schottky characteristics were observed in both the samples.

Figure 2 shows the magnetic field dependence for the magnetic moment of the Pd/SrTiO$_{3-x}$/Ti/Au with a 3.57 nm film thickness (Sample A) under an applied reverse bias voltage of -10 V and 0 V. The nonlinear component of the magnetization was observed for both voltages. We confirmed that the magnetic field dependent magnetic moment of SrTiO$_{3-x}$/Ti/Au without Pd as a background measurement does not show the nonlinear components of magnetization. This indicates that the ferromagnetic moment of Pd(100) ultrathin films intrinsically changes when applying a bias voltage.

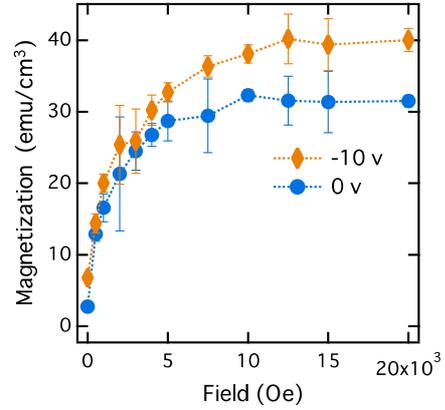

Fig. 2. The nonlinear components of magnetization of the Sample A at bias voltages of -10 V and 0 V with an in-plane magnetic field. Saturation magnetization was different depending on the voltage.

As shown in Fig. 2, the magnetization of Pd was saturated over 12,500 Oe. Then, we measured the magnetic moment of Samples A and B at 12,500 Oe while changing the bias voltage from -10 V to 0 V, as shown in Fig. 3. In Sample A, the magnetic moment monotonically increased as the bias voltage increased. In Sample B, the magnetic moment initially decreased as the bias voltage increased and then increased after passing a minimum value at approximately -8 V. Thus, the voltage dependent magnetic moment depends on the Pd film thickness. The measurements were repeated to verify the reproducibility of the results, and the magnetic moment behavior was reproduced in both the samples. However, changes in the magnetic moment were noticeable when increasing the number of measurements. Such a voltage dependence of the magnetic moment was observed only in the samples exhibiting Schottky characteristics. The observed non-monotonic behavior depends on the Pd film thickness, which is not expected from the parasitic effects of Joule heating or from magnetic fields generated by a leakage current. Therefore, the voltage dependence of the magnetic moment is attributed to changes in the magnetic moment due to modulations of the QWs. We consider the observed behavior in terms of the phase accumulation model.

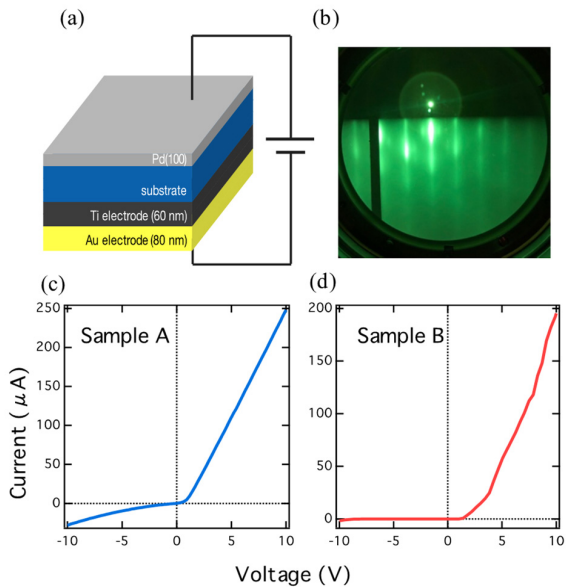

Fig. 1. (a) Schematic illustration of the sample and (b) the RHEED image of the Pd(100)/STO/Ti/Au sample. The observed streak lines indicate high crystallinity and atomic flatness and the *I-V* curves of (c) Sample A and (d) Sample B.

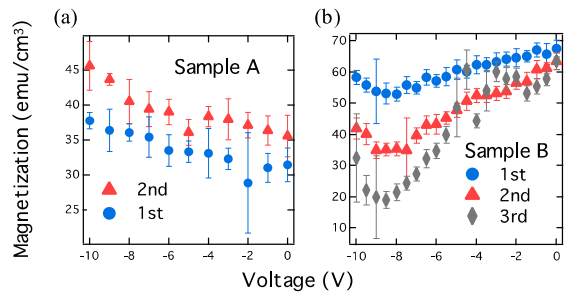

Fig. 3. Voltage dependence of the magnetic moment of (a) Sample A and (b) Sample B at 12500 Oe.

The Pd(100) film exhibits ferromagnetism when the $k_z$ matches the Fermi wave number $k_F$ at 0.83.[21] In other words, the condition under which ferromagnetism appears in the Pd(100) film is determined by $N$, $\Phi$, and $j$ according to Eq. (2). Thus, changes in $\Phi$ at the metal/semiconductor interface from an applied bias voltage to the Schottky barrier[22] should modify the magnetism of the Pd(100) films. As shown in Ref. 6, the magnetization of the Pd(100) films is a periodic function with respect to $N$. Assuming that $\Phi$ decreases when applying a bias voltage, the associated value of $N$ increases to maintain the condition $k = k_F$.

In the schematic diagram in Fig. 4, the solid and dashed lines show the periodic, $N$-dependent magnetizations corresponding to the two amplitudes of $\Phi$ with and without a bias voltage, respectively. As seen the figure, a magnetic state transitions to another state corresponding to a smaller $N$ value when a bias voltage is applied. We interpret the magnetic moment of Samples A and B depending on the bias voltage based on the results of Fig. 4. This figure reproduces the film thickness dependence of the magnetization obtained from a previous experiment in Ref. 6, which shows the film thicknesses of Samples A and B. As a bias voltage is applied, Sample A shows a monotonic increase in the magnetic moment, while Sample B shows the magnetic moment has a minimum, which is consistent with Fig. 3.

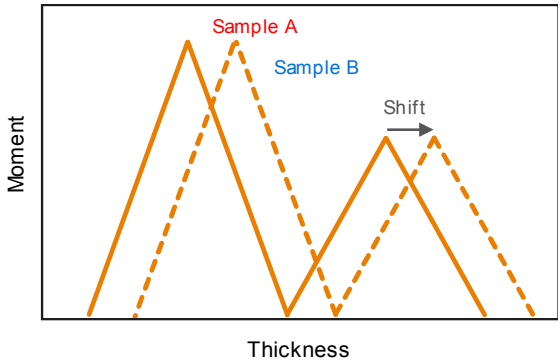

Fig. 4. Schematic image of the $N$-dependent magnetizations corresponding to the two amplitudes of $\Phi$ with and without a bias voltage. As $\Phi$ decreases due to the applied voltage, the relationship shifts to the right. Thus, the magnetization of the Pd(100) ultrathin films changed along with this shift.

Next, $\Delta\Phi$ was evaluated, which is the amount of change in $\Phi$ that accompanies the applied bias voltage. To perform the evaluation, we consider the bias voltage dependent magnetic moment of Sample B with a minimum occurring at a voltage of -8 V. In a previous study,[6] the magnetization of Pd(100) ultrathin films showed a minimum at a thickness of 3.80 nm ($N$ = 19.5). This represents a difference from Sample B by 0.07 nm, or 0.3 atomic layers. This indicates that the magnetic state shifts corresponding to a film thickness as thin as 0.3 atomic layers by applying the bias voltage of -8 V. Substituting this shift into Eq. (1) and assuming that the quantum number $j$ does not change, a $\Delta\Phi$ of $0.43\pi$ is found. This is comparable to a $\Delta\Phi$ of $0.5\pi$, which was previously observed when a bias voltage was applied to a Pb/Si Schottky barrier.[22]

It was confirmed that the change in the observed magnetic moment from an applied bias voltage is caused by $\Delta\Phi$, as opposed to contributions from the charging effect. To this end, a Pd thin film of 3.29 nm thick was deposited on a STO substrate having no semiconductor characteristics, which was prepared at a low annealing temperature of 680 ℃ (Sample C). In Sample C, which shows no Schottky characteristics, as shown in Fig. 5(a), there was minimal observed bias voltage dependence of the magnetic moment, as seen in Fig. 5(b). This indicates that magnetic changes in the Pd(100) ultrathin film due charging effects is negligible under the experimental conditions. A previous study using first-principles calculations on Pd(100) ultrathin films showed that a charge accumulation of $10^{13}$ cm$^{-2}$ or more was required to change the magnetic moment.[24] We evaluate an accumulated $10^{10}$ cm$^{-2}$ charge at an applied bias of 10 V in Sample C by assuming a parallel plate capacitor and using a dielectric constant of 310 at a thickness of 0.5 mm for the STO substrate. Therefore, the charging effect on the magnetism of a Pd(100) ultrathin film is negligible from a theoretical perspective.

Next, we discuss why the bias dependent change in the magnetic moment increases when more measurements are performed. A possible origin is the time dependent change in the redistribution of oxygen vacancies that are formed in the STO. The distribution of oxygen vacancies is non-uniform,[25] causing the Schottky junction to be non-uniformly formed at the interface. In the region with significant oxygen vacancies at the interface, an ohmic junction is formed and changes in the magnetic moment are not expected when applying a bias voltage. The redistribution of oxygen vacancies can occur with an applied voltage because they are positively charged and are carried by the electric field.[25,26] Such a redistribution occurs on the time scales of minutes,[26] which gradually grows the region containing the Schottky junction. This causes a significant change in the bias dependence of the magnetic moment over time.

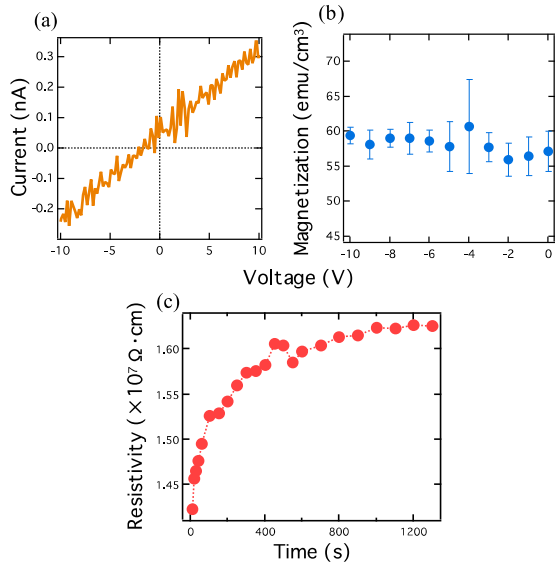

Fig. 5. (a) *I-V* curve and (b) bias voltage dependent magnetic moment of Sample C. Minimal changes in the magnetic moment on the insulator substrate were seen over the considered voltage range. (c) Time dependent electrical resistivity of Pd/SrTiO$_{3-x}$ at a bias voltage of -10 V. The resistivity gradually increased up to approximately 1,000 seconds.

Figure 5(c) shows the time dependent electrical resistivity at a bias voltage of -10 V. The resulting gradual change in the resistivity is observed up to approximately 1,000 seconds. This suggests that the resistivity change is caused by moving oxygen vacancies in the electric field at the interface. Therefore, we believe that changes in the magnetic moment that become noticeable over time are attributed to the growth of the Schottky junction from the movement of oxygen vacancies.

This study shows that the saturation magnetization of Pd(100) ultrathin films changes when applying a bias voltage to the Schottky barrier. The change in the magnetic moment can be interpreted through the modulation of the phase shift at the interface based on the phase accumulation model. Elucidation of the magnetic manipulation mechanism is a significant task to realize magnetic control using an external field, and this study is expected to be a basis for future work. It is expected that the manipulation of QWs will be applied to control the physical properties of films over wide application areas beyond magnetic control.

This work was supported by JSPS KAKENHI Grant Numbers #15J00298 and 15H01998. One of the authors (S.S.) also acknowledges a fellowship from the JSPS.